\documentstyle[12pt]{article}

\newcommand{\vp}{\varphi}
\newcommand{\ra}{\rightarrow}

\newcommand{\C}{\Gamma}
\newcommand{\Ck}{\Gamma_\kappa}

\newcommand{\be}{\begin{equation}}
\newcommand{\ee}{\end{equation}}
\newcommand{\bea}{\begin{eqnarray}}
\newcommand{\eea}{\end{eqnarray}}
\newcommand{\ba}{\begin{array}}

\newcommand{\ea}{\end{array}}

\def\bbox{{\,\lower0.9pt\vbox{\hrule \hbox{\vrule height 0.2 cm
\hskip 0.2 cm \vrule height 0.2 cm}\hrule}\,}}
\newcommand{\dsl}{\pa \kern-0.5em /}

\newcommand{\ep}{\epsilon}

\begin{document}


\begin{titlepage}
\vfill
\begin{flushright}
hep-th/0002086\\
\end{flushright}

\vfill

\begin{center}
\baselineskip=16pt
{\Large\bf Intersecting Noncommutative D-branes and \\
Baryons in Magnetic Fields}
\vskip 15.mm
{ ~{}Nakwoo Kim
} \\
\vskip 0.5cm
{\small\it
  Department of Physics \\ Queen Mary and Westfield College\\
  Mile End Rd \\ London E1 4NS, UK 
\\ Email : {\tt N.Kim@qmw.ac.uk}
}\\
\vspace{2pt}
\end{center}
\vfill
\begin{center}
{\bf ABSTRACT}
\end{center}
\begin{quote}
We study supersymmetric intersecting configurations of D-branes with
$B$-field backgrounds. Noncommutative D-brane or M-brane pairs can 
intersect supersymmetrically over $(p-1)$-brane, as well as over 
$(p-2)$-brane like ordinary branes. 
$d=10$ and $d=11$ supergravity solutions are obtained 
and the supersymmetry projection rule is examined. As an application 
we study a noncommutative 
D7-brane probe in noncommutative D3-brane background, intersecting
at noncommutative plane, which describes BPS baryons of noncommutative
gauge theory in the context of AdS/CFT correspondence.

\vfill
\vskip 5mm
\end{quote}
\end{titlepage}
\setcounter{equation}{0}
\section{Introduction}
The relevance of noncommutative geometry to superstring theory was
first noted by Connes, Douglas and Schwarz \cite{cds}, who showed that the
compactification of Matrix theory gives Yang-Mills theory defined
on noncommutative torus when a constant background three-form field 
potential is turned on. Recently Seiberg
and Witten \cite{sw} showed that Born-Infeld theory on ordinary space with
nonzero gauge background and noncommutative Born-Infeld theory
are equivalent up to field redefinition, and by adopting
different regularization schemes one can derive both theories
from the dynamics of open strings with mixed boundary conditions.  

Following this one of the most important subject is the study of 
supergravity solutions of D-branes with Neveu-Schwarz-Neveu-Schwarz 
(NS-NS) $B$-fields, or when we generalize to M-theory,
M-branes with three-form $C$-field potential backgrounds. 
It is now well-established that
large-$N$ supersymmetric Yang-Mills theories are holographic
descriptions of string theories in $N$ D-brane backgrounds \cite{malda}.
Various supergravity solutions which are dual to noncommutative Yang-Mills
theories are presented in \cite{hi1,maldarusso}. 
They are characterized by a nonzero $B$ on the boundary but reduce to
the usual D$p$-brane solutions at the horizon. In this paper
we will mainly consider only one magnetic component of $B$ which
is nonzero, and then from the Ramond-Ramond (RR)
fields we can see that the supergravity solutions 
describe bound states of D$p$ and D$(p-2)$ branes. 
From now on we will call these 1/2 BPS D-branes with 
nontrivial $B$ noncommutative D-branes.

The next most essential objects in string theory which are 
relevant to noncommutative geometry should be intersecting 
noncommutative branes with nontrivial $B$.
Supergravity solutions for such configurations are obtained
in \cite{costa} using the $SL(2,Z)$ electro-magnetic duality of
M-theory compactifications.
For ordinary brane pairs without $B$-fields the 
necessary condition for supersymmetry is that the number of overall
transverse directions must be a multiple of 4. In particular
a D$p$-brane pair can make BPS intersecting configuration over
D$(p-2)$-brane. In this paper we point out that with nonvanishing
$B$, a D$p$-brane pair can intersect over $(p-1)$ dimensional space, 
preserving 1/4 of the supersymmetries. An important consequence
of this reasoning applies to the study of baryonic branes in
the AdS/CFT correspondence. In $N$ D3-brane background, a D5-brane plays 
a role of the source for $N$ fundamental strings which constitute
a baryon in the dual theory. We can study the baryons in terms of a soliton of 
D5-brane Born-Infeld action, when the D5-brane is wrapped on $S^5$ part
of ${\rm AdS}_5 \times S^5$. Now about the noncommutative version 
of AdS/CFT correspondence, it turns out that a single D5-brane in 
noncommutative D3-brane background cannot 
enjoy BPS configuration. But we will see that a D7-brane with 
appropriate Born-Infeld gauge field excitations can.
The baryonic D5-brane wrapping $S^5$ must be modified to a D7-brane
which also wraps the $S^5$ and extends on the noncommutative part
of the D3 worldvolume.

The plan of this paper is as follows.  
In Section 2 starting with the supergravity solution for 
noncommutative branes we present various BPS intersecting 
noncommutative brane 
configurations. Section 3 discusses the supersymmetries of 
intersecting noncommutative branes using both the supergravity solution
and the $\kappa$-symmetry of D-brane probe embedded in the
supergravity backgrounds. In Section 4 we find the equation 
governing the shape of baryonic noncommutative D7-brane from the 
Dirac-Born-Infeld 
action. The last section provides a brief discussion.  
\section{Supergravity Solutions}
The supergravity solution of a D3-brane in a constant NS-NS
$B$ field background is obtained in \cite{hi1,maldarusso}. 
\bea \label{ncd3g}
&& ds^2_{\rm string} = f^{-1/2} [ -dx^2_0  + dx^2_1 +
h ( dx^2_2 + dx^2_3 ) ] + f^{1/2} ( dr^2 + r^2 d\Omega^2_5 ), 
\nonumber \\
&& f = 1 + \frac{\alpha'^2 R^4 }{r^4}, \quad\quad h^{-1} = \sin^2 \vp
 f^{-1} + \cos^2 \vp , 
\nonumber \\
&& B^{\rm NS}_{23} = \tan \vp f^{-1} h , \quad \quad e^{2 \phi} = g^2 h , 
\nonumber \\
&& F^{\rm RR}_{01r} = \frac{1}{g} \sin \vp \partial_r f^{-1},
\quad \quad
F^{\rm RR}_{0123r} = \frac{1}{g} \cos \vp h \partial_r f^{-1} .
\eea
Note that $B_{23}= 0$ close to the horizon, while $B_{23} = 
\tan \vp$ on the boundary. From the RR gauge fields it is evident 
that this solution has a D-string charge as well as a D3-brane.
Technically there are two ways to get above solution \cite{hi1,bmm}.
We can either (1) start
with a D1-brane solution with a constant $B^{\rm NS}_{23}$ and delocalized
in $23$-directions, and then
T-dualize twice, along $x_2$ and $x_3$, or (2)
consider a D2-brane, extended along $12$ and delocalized 
in $x_3$-direction, rotate the solution in $23$-plane and then T-dualize
along $x_3$. These two methods produce gauge equivalent solutions
with different values of $B^{\rm NS}_{23}$. Method (1) gives above 
solution and method (2) gives the same solution with
$B^{(2)}_{23} = B^{(1)}_{23} - \tan \vp$, which means we exchange
two solutions at the expense of turning on a U(1) Born-Infeld
field strength on the D3-brane worldvolume. In the context of 
holography $r/\alpha'$ represents energy scale of the boundary theory.
The fact that $B^{\rm NS} = 0$ at $r=0$ is consistent with the
classical expectation that the noncommutativity has no effect on
IR behaviour. We should comment here that this is usually not true
in quantum field theories defined on noncommutative spaces. It turns out that
in loop integrals the high momentum modes can generate long range 
forces and there is a mixing of IR and UV physics in noncommutative 
field theory \cite{mrs,mst}.

To decouple the aymptotic region and relate to noncommutative Yang-Mills
theory, we rescale the parameters in the following way 
\cite{maldarusso,aliozjab},
\bea
\label{decouple}
&& \alpha' \ra 0 , \quad\quad \tan \vp = \frac{\Theta}{\alpha'},
\nonumber\\
&& x_{0,1} = \bar{x}_{0,1}, \quad\quad 
x_{2,3} = \frac{\alpha'}{\Theta}\bar{x}_{2,3} ,
\\
&& r = \alpha' u , \quad\quad g = \alpha' \bar{g},
\nonumber
\eea
where $\Theta,u,\bar{g},\bar{x}_i$ are fixed. Especially $\Theta$ is the
noncommutativity parameter, $[\bar{x}_2,\bar{x}_3 ] \sim \Theta$.
Several basic aspects of noncommutative AdS/CFT correspondence, like
the calculation of correlation functions and Wilson loops, were
studied using the above supergravity solution and decoupling limit
\cite{maldarusso,aliozjab}.

Now we turn to intersecting D-brane solutions with nontrivial $B$-fields.
In \cite{costa} the solutions were introduced as non-threshold
bound states of orthogonally intersecting branes while in this
paper we rather call them intersecting noncommutative D-branes.
We use here the method (2) explained above, i.e. T-dualizing D-branes 
at an angle.

Let us start with a D4-brane pair intersecting over
a D2-brane with $B$ fields. To obtain the supergravity solution we first 
a write Type IIB solution describing a D3-brane pair intersecting
at a D1-brane, which can be written easily using the
harmonic superposition rule,
\bea \label{d3d3at1}
&& ds^2_{\rm string} = f^{-1/2}_1 f^{-1/2}_2 ( -dx^2_0 + dx^2_3)
  + f^{-1/2}_1 f^{1/2}_2 ( dx^2_1 + dx^2_2 ) 
\nonumber\\ 
&& \quad \quad + f^{1/2}_1 f^{-1/2}_2 ( dx^2_5 + dx^2_6 ) 
+ f^{1/2}_1 f^{1/2}_2  ( dx^2_4 + dr^2 + r^2 d\Omega^2_2 ), 
\nonumber \\
&& F_{0123r} = \frac{1}{g} \partial_r f^{-1}_1,
\quad \quad
F_{0356r} = \frac{1}{g} \partial_r f^{-1}_2,
\quad \quad f_{1,2} = 1 + \frac{\alpha'^{1/2} R_{1,2} }{r}. 
\eea
Note that we prepared ${\rm D}3(123) \perp {\rm D}3(356)$
delocalized in $x^4$.
We rotate above solution by introducing new coordinates,
\bea
x_3 &=& \tilde{x}_3 \cos \varphi  - \tilde{x}_4 \sin \varphi ,  \nonumber \\
x_4 &=& \tilde{x}_3 \sin \varphi  + \tilde{x}_4 \cos \varphi .  \nonumber
\eea
And we T-dualize along $\tilde{x}_4$, using the T-duality relation between
supergravity solutions with RR fields \cite{bho} to get the following
solution,
\bea
\label{d4d4-2}
&& ds^2_{\rm string} = f^{-1/2}_1 f^{-1/2}_2 
[ -dx_0^2 + h ( dx^2_3 + dx^2_4 ) ]
+ f^{1/2}_1 f^{-1/2}_2 ( dx^2_1 + dx^2_2 ) 
\nonumber \\
&& \quad\quad + f^{-1/2}_1 f^{1/2}_2 ( dx^2_5 + dx^2_6 )
+ f^{1/2}_1 f^{1/2}_2 (dr^2 + r^2 d \Omega^2_2 ) ,
\nonumber \\
&& e^{2 \phi} = g^2 f^{-1/2}_1 f^{-1/2}_2 h , \quad \quad 
h^{-1} = \cos^2 \vp + f^{-1}_1 f^{-1}_2 \sin^2 \vp ,
\nonumber \\
&& B_{34} = f^{-1}_1 f^{-1}_2 h \tan\vp , 
\nonumber \\
&& F = 
\frac{\cos\vp}{g} R_1 dx_5 \wedge dx_6 \wedge \epsilon_2 
+ \frac{\cos\vp}{g} R_2 dx_1 \wedge dx_2 \wedge \epsilon_2 
\nonumber \\
&&
+ \frac{\sin\vp}{g} df_1^{-1} dx_0 \wedge dx_1 \wedge dx_2 
+ \frac{\sin\vp}{g} df_2^{-1} dx_0 \wedge dx_5 \wedge dx_6 ,
\eea
where $\epsilon_2$ is the volume element of a unit radius 2-sphere.
Note that we shifted $B_{34}$ by $\tan\vp$ to make it vanish
at infinity and $B_{34}=\tan\vp$ at $r=0$. In the next section it
will be shown that this shift guarantees the supersymmetry of a D4(1234)
and a D4(3456) probe {\it without} Born-Infeld U(1) gauge field in the above 
background.
By varying $\vp$ we note that at $\vp=0$ we have a D4-brane pair of
(1234) and (3456), while at $\vp=\pi/2$ we have a D2-brane pair intersecting
over a point, extended on (12) and (34) plane respectively.

Now let us consider a noncommutative D3-brane pair intersecting over
a 2d plane. We start with an intersecting D2-brane pair
${\rm D}2(12)\perp{\rm D}2(34)$, 
\bea \label{d2d2}
&& ds^2_{\rm string} = - f^{-1/2}_1 f^{-1/2}_2 dx^2_0  + 
f^{-1/2}_1 f^{1/2}_2 ( dx^2_1 + dx^2_2 ) 
\nonumber\\ && \quad \quad + f^{1/2}_1 f^{-1/2}_2 ( dx^2_3 + dx^2_4 ) 
+ f^{1/2}_1 f^{1/2}_2  ( dr^2 + r^2 d\Omega^2_4 ), 
\nonumber \\
&& f_{1,2} = 1 + \frac{\alpha'^{3/2} R^3_{1,2} }{r^3}, 
\quad \quad   
e^{2 \phi} = g^2 f^{1/2}_1 f^{1/2}_2,
\nonumber \\
&& F_{012r} = \frac{1}{g} \partial_r f^{-1}_1,
\quad \quad
F_{034r} = \frac{1}{g} \partial_r f^{-1}_2 .
\eea
Then rotate the solution in 23-directions like we did above,
and perform T-duality along $\tilde{x}_3$, to get
\bea 
\label{d3d3-2}
&& ds^2_{\rm string} = -f^{-1/2}_1 f^{-1/2}_2 dx_0^2
+ f^{-1/2}_1 f^{1/2}_2 dx^2_1 
+ f^{-1/2}_1 f^{-1/2}_2 k ( dx^2_2 + dx^2_3 ),
\nonumber \\
&& \quad\quad\quad + f^{1/2}_1 f^{-1/2}_2 dx^2_4 
+ f^{1/2}_1 f^{1/2}_2 (dr^2 + r^2 d \Omega^2_4 ) , 
\nonumber \\
&& e^{2 \phi} = g^2 k , \quad \quad \quad \quad 
k^{-1} = f^{-1}_1 \cos^2 \vp + f^{-1}_2 \sin^2 \vp ,
\nonumber \\
&& B_{23} = -\tan\vp f^{-1}_1 k , 
\nonumber \\
&& F_{01r} = \frac{1}{g} \cos \vp \partial_r f^{-1}_1,
\quad\quad   F_{04r} = -\frac{1}{g} \sin \vp \partial_r f^{-1}_2 ,
\nonumber \\
&& F_{0123r} = \frac{1}{g} \sin \vp k f^{-1}_2 \partial_r f^{-1}_1,
\quad\quad   F_{0234r} = \frac{1}{g} \cos \vp k f^{-1}_1 \partial_r f^{-1}_2 .
\eea
At $\vp=0$ or $\vp=\pi/2$ we see that this system reduces to
D3-brane and D1-brane intersecting at a point, but at generic
values we have a pair of D3-branes with nontrivial $B$-fields.
We also shifted the value of $B$ by $\tan\vp$ to make $B=\tan\vp$
on the boundary, but differently from previous examples it is nonzero 
at the horizon. In this background D3-brane probes extended along
(123) and (234)-directions cannot be supersymmetric with the same values
of ${\cal F}=F-B$. In the above background, a D3-brane on (123)-space
is supersymmetric with ${\cal F}_{23}=-\cot\vp f^{-1}_1 k$, while a
D3-brane on (234)-space becomes supersymmetric with ${\cal F}_{23}=
\tan\vp f^{-1}_2 k$. This will be checked in the next section using
the $\kappa$-supersymmetry of D-brane action.

Like ordinary intersecting brane solutions without $B$-fields,
we can obtain other solutions via T-duality. For example if we
T-dualize above solution along $x_5$, we get a noncommutative D4-brane pair
intersecting over 3-brane, i.e. D$4(1235)\perp {\rm D}4(2345)$, with
$B_{23}$. 

M-theory generalization is also easily achieved using the relation 
between IIA string theory and M-theory, which can be found for example
in \cite{bho}. In this process both NS-NS 2-form $B$-field and 
RR 3-form field are united into M-theory 3-form field potential
$C$. A noncommutative M5-brane solution M5(12345) with $C_{012}$ and
$C_{345}$, which can be considered as a bound state of
M$5(12345)$ and M$2(12)$ can be written easily from the noncommutative 
D4-brane on (1234)-space with $B_{34}$,
\bea
&& ds^2_{11} = f^{-1/3} h^{-1/3}
[ -dx_0^2 + dx^2_1 + dx_2^2 + h
( dx^2_3 + dx^2_4 + dx^2_5 ) + f (dr^2 + r^2 d\Omega^2_4) ],
\nonumber \\
&& f = 1 + \frac{R^3}{r^3} , \quad\quad\quad 
h^{-1} = f^{-1} \sin^2 \vp + \cos^2 \vp , 
\nonumber \\
&& dC_3 = \sin \vp \; df^{-1} \wedge dx_0 \wedge dx_1 \wedge dx_2
 - 6 \tan \vp \; d(f^{-1} h) \wedge dx_3 \wedge dx_4 \wedge dx_5
\nonumber \\
&& \quad\quad\quad + \cos \vp \; 3 R^3 \epsilon_4 ,
\eea
where $\epsilon_5$ is the volume element of a unit radius 5-sphere.

For a noncommutative M5-brane pair intersecting over 3d space,
${\rm M5}(12345) \perp {\rm M5}(34567)$ with
$C_{345},C_{012},C_{067}$, the supergravity solution can be 
obtained from the IIA solution eq.(\ref{d4d4-2}),
\bea
&& ds^2 = f^{-1/3}_1 f^{-1/3}_2 h^{-1/3} [ 
- dx^2_0 + f_2 ( dx^2_1 + dx^2_2 ) 
+ h ( dx^2_3 + dx^2_4 + dx^2_5 )
\nonumber \\
&& \quad\quad + f_1 (dx^2_6 + dx^2_7) + f_1 f_2 (dr^2 + r^2 d\Omega_2^2 )
] ,
\nonumber \\
&& f_{1,2} = 1 + \frac{R_{1,2}}{r} , \quad \quad 
h^{-1} = \cos^2 \vp + f^{-1}_1 f^{-1}_2 \sin^2 \vp , 
\nonumber \\
&& dC_3 = \sin\vp \; df^{-1}_1 \wedge dx_0 \wedge dx_1 \wedge dx_2
+ \sin\vp \; df^{-1}_2 \wedge dx_0 \wedge dx_6 \wedge dx_7
\nonumber \\
&& \quad\quad\quad + R_1 \cos\vp \; dx_6 \wedge dx_7 \wedge \ep_2
+ R_2 \cos\vp \; dx_1 \wedge dx_2 \wedge \ep_2
\nonumber \\
&& \quad\quad\quad - 6 \tan\vp \; d(f^{-1}_1 f^{-1}_2 h ) 
\wedge dx_3 \wedge dx_4 \wedge dx_5 .
\eea

A noncommutative M5-brane pair can also intersect over 4d space,
${\rm M5}(12346) \perp {\rm M5}(23456)$ with 
$C_{234},C_{015},C_{016}$. The D=11 supergravity condition can be 
obtained by lifting eq.(\ref{d3d3-2}) twice,
\bea
&& ds^2 = k^{-1/3} f^{-1/3}_1 f^{-1/3} [
(-dx^2_0 + dx^2_1)
+ k ( dx^2_2 + dx^2_3 + dx^2_4 )
\nonumber \\
&& \quad\quad 
+ f^{-1}_2 dx^2_5 + f^{-1}_1 dx^2_6 
+ f_1 f_2 (dr^2 + r^2 d\Omega^2_3 )] , 
\nonumber \\
&& f_{1,2} = 1 + \frac{R^2_{1,2}}{r^2}
, \quad\quad
k^{-1} = f^{-1}_1 \cos^2 \vp + f^{-1}_2 \sin^2 \vp,
\nonumber \\
&& dC_3 = 
\cos\vp \; df^{-1}_1 \wedge dx_0 \wedge dx_1 \wedge dx_5
- \sin\vp \; df^{-1}_2 \wedge dx_0 \wedge dx_5 \wedge dx_6
\nonumber \\
&& \quad\quad\;
+ 2 R_1^2 \sin\vp\; dx_6 \wedge \epsilon_3 
+ 2 R_2^2 \cos\vp\; dx_5 \wedge \epsilon_3
\nonumber\\
&& \quad\quad\; +6 \tan\vp \; d( f^{-1}_2 k )\wedge
dx_2 \wedge dx_3 \wedge dx_4 .
\eea
\section{Supersymmmetry of the solutions}
\subsection{Killing spinor equations of supergravity}
The solutions we have considered in the last section are all supersymmetric,
in particular they preserve 1/2 or 1/4 of the total 32 supersymmetries.
The preserved supersymmetry can be studied using the supersymmetry
transformations of the bosonic fields of the supergravity theory.
For IIB string we may consider the supersymmetry variations of the
dilatino and the gravitino, which are written as follows, in Einstein frame,
\bea
\delta \lambda &=&
\frac{-1}{2 \tau_2} 
( \frac{\tau^* - i}{\tau + i} )
\Gamma^M \partial_M \tau (\eta_1 - i \eta_2 )
- \frac{i}{24} \Gamma^{MNP} G_{MNP} (\eta_1 + i \eta_2 ) ,
\nonumber\\
\delta \psi_M &=&
\partial_M 
(\eta_1 + i \eta_2 ) 
+ \frac{1}{4} \omega^{ab}_M \Gamma^{ab}
(\eta_1 + i \eta_2 ) 
+\frac{1}{8\tau_2} 
[ ( \frac{\tau - i}{\tau^* - i} ) + {\rm c.c.} ]
(\eta_1 + i \eta_2 ) 
\nonumber \\
&& + \frac{i}{480} \Gamma^{M_1 \cdots M_5}\Gamma_M F_{M_1 \cdots M_5}
(\eta_1 + i \eta_2 ) 
\nonumber \\
&& - \frac{i}{96} ( \Gamma^{NPQ}_M G_{NPQ} - 9 \Gamma^{NP} G_{MNP} )
(\eta_2 + i \eta_1 ) ,
\eea
where
\be
\tau = \tau_1 + i \tau_2 = \tau_1 + i e^{-\phi},
\ee
\be
G_{MNP} = i \sqrt{\tau_2} \frac{|1 - i \tau|}{\tau_2 (1 - i \tau)} 
(F^{\rm RR} - \tau F^{NS} )_{MNP}.
\ee
Substituting the supergravity solutions for IIB D$p$-branes
we obtain the BPS conditions,
\be
( \sigma_3 )^{\frac{p-3}{2}} i \sigma_2 
\otimes \bar{\Gamma}_{01\cdots p}
\epsilon = \epsilon ,
\ee
with
$$ \epsilon = \pmatrix{\eta_1 \cr \eta_2}, $$
where $\eta_{1,2}$ are two left-handed Majorana-Weyl spinors.
$\bar{\C}_M$ represent flat space gamma matrices.
It is useful to remember that 
for F-strings the supersymmetry projection is $\sigma_3 
\bar{\Gamma}_{01} \epsilon = \epsilon$. 

For a D3-brane with $B$-field, the Killing spinor equation 
of the solution eq.(\ref{ncd3g}) gives
\be
\label{ncd3k}
h^{1/2} ( i\sigma_2 \bar{\Gamma}_{0123} \cos \vp
- f^{-1/2} \sigma_1 \bar{\Gamma}_{01} \sin \vp ) \epsilon = \epsilon .
\ee
$\vp \rightarrow 0 $ is the limit of ordinary D3-brane while
$\vp \ra \pi /2$ is relevant to the low energy noncommutative 
Yang-Mills theory as can be seen from eq.(\ref{decouple}). 
We notice that the supersymmetry projection rule
is the same as a D1-brane along 1-direction in that case.

For a D4-brane with $B_{34} \neq 0$, which can be obtained from
eq.(\ref{ncd3g}) via T-duality, similarly we have 
\be
h^{1/2} ( \Gamma_{11} \bar{\Gamma}_{01234} \cos \vp
- f^{-1/2} \bar{\Gamma}_{012} \sin \vp ) \epsilon = \epsilon ,
\ee
where $\epsilon$ for IIA theory is a Weyl spinor in 10D.
Again at $\vp=0$ it is the same as ordinary D4-brane, while
at $\vp=\pi/2$ the projection condition becomes that of a D2-brane.
along 12-directions.

Now let us consider intersecting cases. For intersecting pairs
we have a system of BPS conditions, $(\C_{(1)}-1)\epsilon=
(\C_{(2)}-1)\epsilon=0$, with $\C_{(1)}^2=\C_{(2)}^2=1,\;
[\C_{(1)} , \C_{(2)} ] = 0 $, thus preserving 1/4 of the supersymmetries.
If we substitute
the solution of a noncommutative D4-brane pair intersecting over a 2d plane, 
eq.(\ref{d4d4-2}),
into the supersymmetry transformation rule of IIA supergravity, we get
the following set of conditions,
\bea
\label{d4d4k}
\Gamma_{(1)} &=& h^{1/2} ( \cos \vp \Gamma_{11} \bar{\Gamma}_{01234} 
- f^{-1/2}_1 f^{-1/2}_2 \sin\vp\bar{\Gamma}_{012} ) ,
\nonumber \\
\Gamma_{(2)} &=& h^{1/2} ( \cos \vp \Gamma_{11} \bar{\Gamma}_{03456} 
- f^{-1/2}_1 f^{-1/2}_2 \sin\vp\bar{\Gamma}_{056} ) .
\eea
For a noncommutative D3-brane pair solution eq.(\ref{d3d3-2}), 
sharing a 2d-plane, we have
\bea \label{d3d3k}
\C_{(1)} &=& k^{1/2} ( \sin \vp f^{-1/2}_2 i\sigma_2 \bar{\Gamma}_{0123} 
+ \cos \vp f^{-1/2}_1 \sigma _1 \bar{\Gamma}_{01} ) ,
\nonumber \\
\C_{(2)} &=& k^{1/2} ( \cos \vp f^{-1/2}_1 i \sigma_2 \bar{\Gamma}_{0234} 
- \sin \vp f^{-1/2}_2 \sigma_1 \bar{\Gamma}_{04} ) .
\eea
\subsection{$\kappa$-symmetry and D-brane probes}
Alternatively one can use $\kappa$-symmetry of D-brane action
to study the supersymmetry of intersecting D-brane configurations.
$\kappa$-symmetry is a fermionic gauge symmetry on the worldvolume,
which produces a global worldvolume supersymmetry when combined
with the global target space supersymmetry upon gauge fixing.
Thus $\kappa$-symmetry plays an essential role in formulating
supersymmetric D-brane actions \cite{dbrane}. And it is also useful
in studying supersymmetrically intersecting configuration of D-branes 
\cite{bdl,balal,bkop}.
For a brane probe configurations the fraction of preserved 
supersymmetry is determined by the following equation combined with
the supersymmetry breaking condition of the gravity background,
\be
(1 - \Ck ) \epsilon = 0,
\ee
where $\epsilon$ is the spacetime supersymmetry parameter, and
$\Ck$ is an Hermitian traceless matrix, satisfying
\be
{\rm tr} \Ck = 0 , \quad \quad \Ck^2 = 1.
\ee
$\Ck$ is nonlinear in ${\cal F}=F-B$, where $F$ is the Born-Infeld
2-form field strength and $B$ is the pull-back of the NS-NS two-form
gauge potential. 
The explicit form will be important for our analysis.
\be \label{kappa}
\Ck = \frac{\sqrt{|g|}}{\sqrt{|g+{\cal F}|}}
\sum^{\infty}_{n=0} \frac{1}{2^n n!} \gamma^{\mu_1\mu_1 \ldots \mu_n\nu_n}
{\cal F}_{\mu_1\nu_1} \ldots {\cal F}_{\mu_n\nu_n}
J^{(n)}_{(p)} ,
\ee
where $g$ the induced metric by the map $X$, and 
\be
J^{(n)}_{(p)} = \cases{  (\Gamma_{11})^{n+\frac{p-2}{2}} \Gamma_{(0)} & IIA,
\cr
(-1)^n (\sigma_3)^{n+\frac{p-3}{2}}i \sigma_2 \otimes \C_{(0)} & IIB, }
\ee
and 
\be
\C_{(0)} = \frac{1}{(p+1)! \sqrt{|g|} } \epsilon^{i_1 \cdots
i_{(p+1)}} \gamma_{\mu_1 \cdots \mu_{(p+1)}} .
\ee
The matrix $\gamma_{\mu_1 \cdots \mu_{(p+1)}}$ is the antisymmetrized product
of the worldvolume gamma matrices $\gamma_{\mu}$, 
\be
\gamma_\mu = \partial_\mu X^M \C_M ,
\ee
where $\C_M$ are the spacetime gamma matrices.

Let us first consider $\Ck$ for ordinary D-brane backgrounds. 
Naturally a D$p$-brane probe embedded parallel to the 
background D$p$-branes is supersymmetric, which means
the supersymmetry condition from $\kappa$-symmetry
consideration is identical to the one from Killing spinor
equation of supergravity. For a noncommutative D-brane background
we note that the value of $B$ in the supergravity solution 
eq.(\ref{ncd3g}) is set in the way that a probe D3-brane
{\it without} U(1) Born-Infeld field becomes supersymmetric. For example
we consider a D3-brane in the background of eq.(\ref{ncd3g}),
with worldvolume coordinates $(t,\xi^i) , i=1,2,3$,
$$
X^0 = t , \quad \quad X^i = \xi^i ,
$$
then we get, for ${\cal F}= -B$ i.e. $F=0$,
$$
\Ck = h^{1/2}( i\sigma_2 \bar{\Gamma}_{0123} \cos\vp 
- f^{-1/2} \sigma_1 \bar{\Gamma}_{01} \sin\vp ) ,
$$
which is exactly the same as the BPS condition eq.(\ref{ncd3k}) 
obtained from supergravity.
This is true for noncommutative D$p$-brane pairs intersecting
over $(p-2)$d space. For example
there are two commuting projection operators in eq.(\ref{d4d4k}),
each coinciding with $\Ck$'s for D4-brane lying on 1234-directions
and 3456-directions, both with $F=0$.

The situation is different with the supergravity solution of a 
noncommutative D3 pair intersecting over a plane, eq.(\ref{d3d3-2}). 
The value of $B$ is shifted to give a D3 probe extended
in 234-directions without Born-Infeld field supersymmetric. 
For the other noncommutative D3-brane extended along 123-directions,
the first line of eq.(\ref{d3d3k}) corresponds to the $\Ck$
of a D3-brane probe with ${\cal F}_{23}=\cot 2\vp f^{-1}_1 k$, or
$F_{23} = 2 \csc 2\vp$. In short, eq.(\ref{d3d3-2}) describes a supersymmetric
D3-brane pair intersecting at 23-plane, with different values of 
${\cal F}_{23}$, the difference being $2\csc 2\vp$.

\section{Baryon for noncommutative AdS/CFT}
It was first suggested in \cite{witten} that a D5-brane can be used
as the source of fundamental strings whose end points on D3-branes
are considered as baryon in the context of AdS/CFT correspondence,
and the baryon mass was calculated following this idea in \cite{bamass}.
The fundamental strings in turn can be studied in terms of electrically
charged solitons of D5-brane Born-Infeld action. The supersymmetry
condition for the baryonic D5-brane was first obtained in \cite{ima} 
and the explicit solution
was found and analyzed in detail in \cite{callan}. M-theory
generalization was also achieved in \cite{grst}.

$N$ coincident D3-branes 
at the origin generate the following supergravity solution,
\bea
ds^2 &=& f^{-1/2} ds^2 ({\rm E}^{1,3}) + f^{1/2} ( dr^2
+ r^2 d\Omega^2_5 ) , 
\nonumber\\
G_{(5)} &=& 4 R^4 ( \omega_{(5)} + \star \omega_{(5)} ) ,
\eea
which is $\vp=0$ case of eq.(\ref{ncd3g}), and we use
the near horizon limit $f = R^4/r^4$. The parameter $R$ is
given by $R^4=4\pi g N$. We will study the behaviour
of a D5-brane probe with unit tension, wrapping $S^5$ part of the above 
background. The Hanany-Witten effect \cite{hw} is realized in terms of the 
Wess-Zumino term in the D-brane action. The configuration is represented 
as follows,
\be
\ba{ccccccccccl}
D3 : &1&2&3&-&-&-&-&-&-& 
{\rm background} \nonumber\\
D5 : &-&-&-&4&5&6&7&8&-& 
{\rm probe} \nonumber\\
F1 : &-&-&-&-&-&-&-&-&9& 
{\rm soliton}
\ea
\ee
where 4,5,6,7,8-directions are 5 angles of $S^5$, 9-direction is
the radial direction. The D3-brane supergravity background
gives the projection rule
\be
\label{d3pr}
i\sigma_2 \bar{\C}_{0123} \epsilon = \epsilon ,
\ee
where $\epsilon$ is a covariantly constant spinor of $S^5$.
For our discussion in this section it is sufficient to consider
bosonic sector of D-brane action, 
\be
S = - \int d^6 \xi \left\{ e^{-\phi} \sqrt{-{\rm det} (g+{\cal F}) } 
- V \wedge G_{(5)} \right\} .
\ee
Let $\sigma^\mu = ( t , \theta^i), i=1,...,5$
be the worldvolume coordinates of D5-brane and we choose the static gauge
to fix the worldvolume diffeomorphisms
\be
X^0 = t , \quad\quad X^{i+3} = \theta^i \quad (i=1,2 \cdots 5) .
\ee
And we take $X^1=X^2=X^3=0$, the only activated scalar $X^9=r$.
From the equation $(1-\Ck)\epsilon=0$ and the consistency with
the background eq.(\ref{d3pr}) it can be derived that
the BPS condition is \cite{grst}
\be
F_{0i} = \partial_i (r \cos \theta ) ,
\ee 
where $\theta$ is the polar angle of $S^5$. With this BPS
equation and ansatze of SO(5) symmetry, the Gauss law
leads to
\be
\label{bareq}
\left[ \sin^4 \theta \frac{(r \cos\theta)'}{(r \sin\theta)'}  
\right]' = -4 \sin^4 \theta ,
\ee
which can be solved analytically \cite{callan}
\be
r(\theta) = \frac{A}{\sin\theta} 
\left[ \frac{\eta (\theta)}{\pi (1-\nu)} \right]^{1/3},
\quad\quad \eta (\theta) = \theta - \pi\nu - \sin\theta \cos\theta , 
\ee
where $A$ is an arbitrary scale factor reflecting the conformal 
symmetry and $\nu$ is an integration constant, 
which is related to the number of fundamental strings connecting the D3 and 
D5-brane.

Now let us turn to the noncommutative case. 
The supersymmetry projection rule of the
noncommutative D3-brane background is obtained already. Because of the term 
representing D1-brane charge it is obvious that trivial
D5-brane configuration is not supersymmetric in this 
background. It is still true if we allow for arbitrary
Born-Infeld and scalar excitations on the D5-brane. A hint is given from
the example studied earlier, a noncommutative D3-brane pair intersecting
over 2-plane. Let us consider a D7-brane, with nonzero ${\cal F}_{23}$.
Using the supergravity background eq.(\ref{ncd3g}) and 
expanding eq.(\ref{kappa}), we get
\be \label{ncd7}
\C_{\kappa} = \frac{1}{\sqrt{h^2 + f ({\cal F}_{23})^2 } }
(h \bar{\C}_{02345678} i \sigma_2 + f^{1/2} 
{\cal F}_{23} \bar{\C}_{045678} \sigma_1 ) ,
\ee
where $x_{4,5,6,7,8}$ are angular coordinates of $S^5$.
It is straightforward to check that $\Gamma_{\kappa}$ in eq.(\ref{ncd3k}) 
and eq.(\ref{ncd7}) commute with each other, when
\bea
{\cal F}_{23} &=& h \cot \vp  \nonumber \\
& = & 2 \csc 2\vp - B_{23} .
\eea
So in the supergravity background of noncommutative D3-branes, a
D7-brane along 2345678-directions is supersymmetric when a constant 
Born-Infeld gauge field $2 \csc 2\vp$ is turned on, in addition to the 
NS-NS background. 

We are thus led to consider the Born-Infeld action of D7-brane, using 
the induced metric in the background of eq.(\ref{ncd3g}). It is 
\bea
S &=& -T_7 \int d^8 \xi e^{-\phi} \sqrt{- \det ( g + {\cal F}) } + T_7 
\int d^8 \xi A_\alpha \partial_\beta X^{M_1} \cdots \partial_\gamma
X^{M_7} F_{M_1 \cdots M_7} \nonumber \\
&& + T_7 \int d^8 \xi A_\alpha {\cal F}_{23} \partial_\beta X^{M_1}
\cdots \partial_\gamma X^{M_5} F_{M_1 \cdots M_5} ,
\eea
where $F_{M_1 \cdots M_7}$ is the dual of RR 3-form field strength
in eq.(\ref{ncd3g}), which couples to D1-brane, while $F_{M_1 \cdots M_5}$
is the RR 5-form field strength which couples to D3-brane and is self-dual.

Following \cite{callan} we can proceed in the same way to write 
the action in detail in the near horizon limit, for D7-brane with SO(5)
symmetry. D7 worldvolume coordinates are $\xi^\mu = (t,y,z,\theta^i), 
\; i=1,2...5$
\be
X^0 = t, \quad X^2 = y , \quad X^3 = z , \quad X^{i+3} = \theta^i ,
\ee
and $X^1=0$ and the radial coordinate as the only activated scalar,
$X^9 = r$ as function of $y,z,\theta\equiv\theta^1$.
\be
S = \frac{R^4 T_7 \Omega_4}{\sin \vp} 
\int dt \, dy \, dz \, d \theta \sin^4 \theta 
( -\sqrt{r^2 + r_{\theta}^2 - F_{0\theta}^2 + K \sin^2 \vp } 
+ 4A_0 ) ,
\ee
where
\bea
K &=& (r_y^2 + r_z^2 ) ( r^2 - F_{0\theta}^2) + 2 F_{0\theta} r_\theta
( F_{0y} r_y + F_{0z} r_z ) - (r^2 + r_\theta^2 ) ( F_{0y}^2 +
F_{0z}^2 ) \nonumber \\ 
&& - r^2 f h^{-1} (F_{0z} r_y - F_{0y} r_z )^2 ,
\eea
and $r_y = \partial_y r$, etc.

Without help from BPS conditions solving the equations from above 
action should be very complicated. We can find the BPS equations from
the condition 
$(1-\Ck)\epsilon = 0$, with Born-Infeld gauge potential $A_0$ and 
scalar field $r$ turned on to describe fundamental 
strings attached on D7, coming from the noncommutative D3-branes lying 
at the horizon. And it has to satisfy 
the condition $[ \C_{\rm sugra} , \C_{\kappa} ] = 0$.
It turns out that the BPS condition is again the same form,
\be
F_{0i} = \partial_i ( r \cos \theta ).
\ee
the only difference being now $i$ includes $y,z$.
$K$ vanishes when we use this BPS equation, so the 
equation of motion is simplified substantially.
Employing the decoupling limit eq.(\ref{decouple}), we finally get
\be
\label{npde}
\sin^{-4} \theta \partial_\theta 
\left( \sin^4 \theta \frac{\partial_\theta ( u \cos \theta )}
{\partial_\theta ( u \sin \theta ) } \right) + \frac{\Theta^2}{2} 
(\partial_{\bar{y}}^2 + \partial_{\bar{z}}^2 ) u^2 = -4 ,
\ee
where $\bar{y},\bar{z}$ represent the coordinates which are noncommutative 
in the dual large $N$ Yang-Mills theory satisfying
$[\bar{y},\bar{z}] \sim \Theta$. At $\Theta=0$ we reproduce 
eq.(\ref{bareq}) but for $\Theta \neq 0$ this is a nonlinear partial 
differential equation, and the usual technique of separation of variables
is not useful in this case. 
The dependence of $u$ on $\theta$ encodes
the flavor SU(4) structure of the baryon, while the dependence on $\bar{y},
\bar{z}$ should describe the size of baryon in the noncommutative plane 
at a specific energy scale
$u$. It is easily seen that if 
$u(\theta,\bar{y},\bar{z};\Theta)$ is a solution to the equation,
$u(\theta,k\bar{y},k\bar{z};\Theta /k)$ is also a solution, which 
implies the size of the baryon in $\bar{y}\bar{z}$-plane scales as 
$\Theta u$. This is consistent
with the philosophy of noncommutative field theory introducing $\Theta$
as a parameter of nonlocality.

\section{Discussion}
In this paper we have studied intersecting noncommutative $p$-branes
which are BPS. The lesson is that there exist orthogonally intersecting
D-brane pairs which are supersymmetric, with the number of total
transverse directions 2 mod 4, as well as 0 mod 4.
The usual harmonic superposition rule 
cannot be applied directly to get the supergravity solutions, but 
we can easily obtain the solutions using tilted D-branes via T-duality. 
Here we considered solutions with only one component of $B$-field 
which is pure magnetic, but we can also consider $B$-fields with larger 
ranks, and $B_{0i} \neq 0$ cases, which generate electric ${\cal F}$. 
The gauge bundle on the D-brane worldvolume then represent a bound state
of D-brane and fundamental strings. Especially supergravity solutions
of D3-branes with self-dual B-fields are obtained in \cite{drt}.
Supersymmetric configurations of D-branes with nonzero $F$ in 
supergravity backgrounds with $B$ are analyzed systematically using 
$\kappa$-symmetry and applied to study the nonlinear deformations of
the instanton equations in \cite{mmms}.

For baryonic branes, it will be interesting to extend this study
to M-theory, following \cite{grst}. 
Because of the self-duality of $C$-field on M5-brane
worldvolume, to get supersymmetric configuration we need to consider 
M9-branes. There are suggestions on M9-brane action \cite{m9} as a 
supersymmetric
nonlinear sigma model. The lack of $\kappa$-symmetry thereof makes it 
difficult to find the supersymmetric configuration. 
We leave the subject of solving the equation 
eq.(\ref{npde}) and investigating baryonic M9-brane for future work. 
\section*{Acknowledgments}
\noindent
This work is supported in part by PPARC through SPG \#613.
I would like to thank J.P. Gauntlett and C.M. Hull for useful 
discussions, S.-J. Rey for correspondence, and especially M.S. Costa for 
bringing my attention to his works.

\end{document}